\documentclass[submit]{aiaa-tcc}	

\usepackage{amsmath,amsthm,amssymb,amsfonts}
\usepackage{algpseudocode}
\usepackage{algorithm}
\usepackage{graphicx}
\usepackage{float}
\usepackage{caption}
\usepackage{subcaption}
\usepackage{color}
\usepackage{cite}
\usepackage{comment} 
\usepackage{setspace}

\usepackage{tabularx}

\usepackage{soul}

\theoremstyle{definition}

\renewcommand{\Re}{\mathbb{R}}

\newcommand{\vk}{{\bf k}}
\newcommand{\vx}{{\bf x}}

\newcommand{\vr}{{\bf r}}

\newcommand{\vu}{{\bf u}}
\newcommand{\vv}{{\bf w}}
\newcommand{\vw}{{\bf w}}

\newcommand{\vl}{{\bf l}}

\newcommand{\vK}{{\bf K}}

\newcommand{\argmax}{\operatornamewithlimits{argmax}}

\newcommand{\T}{{\mbox{\tiny\sf T}}}
\newcommand{\rd}{\textrm{d}}

\newcommand{\tf}{{t_{f}}}

\newcommand{\Est}[1]{\mathbb{E} \left[ #1 \right]}
\newcommand{\gradx}{\nabla_{\vx}}
\newcommand{\gradchi}{\nabla_{\chi}}
\newcommand{\gradu}{\nabla_{\vu}}
\newcommand{\gradv}{\nabla_{\vv}}
\newcommand{\gradxx}{\nabla_{\vx \vx}}
\newcommand{\gradcc}{\nabla_{\chi\chi}}
\newcommand{\graduu}{\nabla_{\vu \vu}}
\newcommand{\gradvv}{\nabla_{\vv \vv}}

\newcommand{\gradux}{\nabla_{\vu \vx}}
\newcommand{\gradvx}{\nabla_{\vv \vx}}
\newcommand{\graduv}{\nabla_{\vu \vv}}

\newcommand{\Var}[1]{\mathrm{Var} \left[ #1 \right]}

\newcommand{\Hinf}{H^\infty}

\theoremstyle{plain}

\theoremstyle{definition}

\theoremstyle{remark}

\renewcommand{\cite}[1]{[\citen{#1}]}

	\title{Learning-Based Nonlinear $\Hinf$ Control via Game-Theoretic Differential Dynamic Programming}

	\author{\normalsize
		Wei Sun\thanks{Assistant Professor, School of Aerospace and Mechanical Engineering, email: wsun@ou.edu. }~~~
		Theodore B. Trafalis\thanks{Professor, School of Industrial and System Engineering, email: ttrafalis@ou.edu.}\\[0pt]
		{\normalsize\itshape
			University of Oklahoma, Norman, OK 73019}
		\\
	}

\begin{document}

	\maketitle

	\begin{abstract}
		In this work, we present a learning-based nonlinear $\Hinf$ control algorithm that guarantee system performance under learned dynamics and disturbance estimate. The Gaussian Process (GP) regression is utilized to update the nominal dynamics of the system and provide disturbance estimate based on data gathered through interaction with the system. A soft-constrained differential game associated with the disturbance attenuation problem in nonlinear $\Hinf$ control is then formulated to obtain the nonlinear $\Hinf$ controller. The differential game is solved through the min-max Game-Theoretic Differential Dynamic Programming (GT-DDP) algorithm in continuous time. Simulation results on a quadcopter system demonstrate the efficiency of the learning-based control algorithm in handling external disturbances. 
		\end{abstract}

	\doublespacing

	\section{Introduction} \label{sec:intro}

	Many of the system models we are dealing with nowadays, whether the model of an aircraft, a manipulator or a biped robot, are only simplifications or approximations of the real systems and are subject to disturbances \cite{book1990modeling, etkin2012dynamics, Freek2012_SequenceManipulation, vukobratovic2012biped}. In the meantime, the performance of model-based controls depends highly on the accuracy of the underlying mathematical model. In order to deal with this problem, we need tools to obtain a dynamical model that takes into account these uncertainties. Machine learning methods such as neural networks \cite{specht1991general,mikolov2010recurrent} have been utilized broadly to deal with uncertainties in approximate models by leveraging measurements to learn the models. However, neural network learning requires extensive parameter tuning and suffers from local minimum problems which reduce the efficiency of the learning process. In recent years, more attention have been drawn to Gaussian Processes (GPs) \cite{williams2006gaussian} for model learning and control. The learned GP models have been applied to data efficient policy search \cite{deisenroth2011pilco,deisenroth2013gaussian}, and combined with dynamic programming to get optimal policy \cite{deisenroth2008approximate,pan2014probabilistic}. It has also been incorporated with robust control \cite{berkenkamp2015safe}, adaptive control \cite{murray2002nonlinear} and model predictive control (MPC) \cite{kocijan2004gaussian} for control optimization. An advantage of GPs that we will exploit in this paper is that they provide
	uncertainty estimates for their predictions, which can in turn be utilized in updating
	the estimated model of the dynamics based on the mismatch between the real dynamics and learned dynamics of the system. 
	
	An efficient control design is required to guarantee its performance after a relatively realistic model is generated. In this paper, we utilize the  $\Hinf$ control to accomplish this goal. The $\Hinf$ control theory \cite{doyle1989state, zhou1996robust,van1992sub,khargonekar1988h, petersen1987disturbance,dorato1967application} aims to achieve robustness of systems against model uncertainty. Specifically, the $\Hinf$ control keeps the sensitivity of the feedback control loop against a disturbance input subject to modeling error such that the disturbance can be suppressed if the gain of the mapping from the state error to the disturbance is bounded in terms of $\Hinf$ norm. 
	It has been shown in \cite{TamerBasar1995} that the $\Hinf$  control problem can be recast as a differential game problem. The objective of the differential game is to obtain a control that minimizes a given performance index under worst possible disturbances or parameter variations. This differential game can be approached through a game-theoretic variation of the differential dynamic programming (DDP)  \cite{jacobson1970} technique. DDP is a trajectory optimization method that iteratively finds  a  locally optimal control policy from some initial control and  state trajectory. 
	Since the introduction of DDP in \cite{jacobson1970}, there has been a large variations of DDP within the controls and robotics communities, including DDP with min-max formulation \cite{MinMaxDDP}, receding horizon DDP \cite{RecedingDDP2007}, and DDP under stochastic dynamics \cite{TheodorouSDDP2010,sun2016stochastic}. As one of the standard methods  for trajectory optimization, DDP also has a broad range of  applications \cite{Todorov2005b,TassaSynthesis_2012,Abbeel07anapplication,Erez_RSS_11,ControlLimited_Yuval_14,Atkeson03nonparametricrepresentation,RandomSamping_Atkesson2008}. In this paper, we present the  Game-Theoretic Differential Dynamic Programming (GT-DDP) algorithm \cite{sun2018min} that deals with the Hamilton-Jacobi-Isaacs (HJI) equation associated with the value function of the differential game problem and find the  $\Hinf$  control of the dynamics learned through GP.

	The rest of the paper is organized as follows. In Section~\ref{sec:learned_dyn} the system model is introduced and the unknown portion of the dynamics is learned through GP. A  soft-constraint differential game associated with the nonlinear $\Hinf$ control design is formulated in Section~\ref{sec:hinf_control}. In Section~\ref{sec:GTDDP} the derivation of the GT-DDP is presented to solve for the differential game problem and obtain the $\Hinf$ control. Simulation results are demonstrated in Section~\ref{sec:Simulation_Results}. Finally, Section~\ref{sec:Conclusion} provides a summary of the results and suggests some possible future research directions.

\section{Learned Dynamics}   \label{sec:learned_dyn}
		\subsection{Problem Formulation}
		
	Consider a dynamical system whose dynamics can be denoted by the summation of two components, a known function obtained from system modeling and an unknown function that represents the model error induced by the disturbance and noise that are not captured by the original model. The continuous-time system dynamics is given by
	\begin{align}
		\frac{\rd \vx(t)}{\rd t}  = \bar{f}(\vx(t), \vu(t)) + \delta f (\vx(t), \vu(t)), \quad \vx(t_0) = \vx_0, \quad t \ge 0, \label{eq:dyn_ori}
	\end{align}
	where
	$\vx(t) \in \Re^n$ is the state of the dynamic system at $t \in [t_0, \tf]$,  $\vu(t) \in U_1 \subset \Re^{m}$ denotes the control of the system where $U_1$ is a convex constraint sets of the control $\vu$, $\bar{f}(\vx(t), \vu(t))$ denotes the nominal nonlinear system model that is known \emph{a priori} and $\delta f (\vx(t), \vu(t))$ denotes the unknown error dynamics. 
	
	The goal of this paper is to find an estimate of the function $\delta f (\vx(t), \vu(t))$ and an associated confidence range of the estimation, which will then be used to design a nonlinear $\Hinf$ controller that drives the system to its desired state despite the uncertainty in the dynamics. 
	
	\subsection{Model Learning}
	The unknown error dynamics  $\delta f (\vx(t), \vu(t))$ can be modeled through a machine learning process. In this subsection, we introduce the Gaussian processes (GP) approach to achieve this task. 
	
	A GP is a stochastic process such that every finite collection of its random variables have a joint Gaussian distribution. It is widely used in nonparametric regression in machine learning to find an approximation of a nonlinear map from an input vector to a function value. Given state-control pairs $\chi = (\vx, \vu)  \in \Re^{m+n}$ at different time instants, and the corresponding function values $\delta f (\chi) = \delta f (\vx, \vu)$,  the error function $\delta f$ can be approximated by a GP defined by a mean function and a covariance function. The latter is also known as the kernel, which serves as a similarity measure of random variables. In this paper, we consider the Gaussian kernel \cite{williams2006gaussian} 
	\begin{align}
	k(\chi_i,\chi_j) = \sigma_s^2 \exp(-\dfrac{1}{2} (\chi_i - \chi_j)^{\T} M (\chi_i - \chi_j)) \label{eq:gaussian_kernel}
	\end{align}
	 as the covariance between two data points, $\delta f (\chi_i)$ and $\delta f (\chi_j)$, where $\chi_i = \chi(t_i)$, $\chi_j = \chi(t_j)$, $i,j\in\mathbb{N}$. In (\ref{eq:gaussian_kernel}), the process variance $\sigma_s^2$ and the scaling matrix $M$ are hyperparameters.
	 
	 Let $X = [\chi_1,\dots,\chi_N]^{\T}$ denote a vector of $N$ state-control pairs and the vector of their corresponding observed error function values is denoted by $\delta f(X) = [\delta f (\chi_1),\dots,\delta f (\chi_N)]^{\T}$, where $\delta f (\chi_i) = \dot{\vx}(t_i) - \bar{f}(\chi_i)$, and let $\chi^*$ be a test input, then the joint probability distribution of the observed output and the predicted function value $\delta f (\chi^*)$ is given by
	 \begin{align}
	 	\begin{bmatrix}
	 	\delta f(X) \\ \delta f (\chi^*) 
	 	\end{bmatrix}
	 	\sim \mathcal{N} \left(0, \begin{bmatrix}
	 	K(X,X) + \sigma_w^2 I &  \vk(X,\chi^*) \\
	 	 \vk(\chi^*,X)  & k(\chi^*,\chi^*)
	 	\end{bmatrix} \right), \label{eq:joint_pd}
	 \end{align}
	 where $K(X,X)$ has entries $K(X,X)_{(i,j)} = k(\chi_i,\chi_j)$, $i,j\in\{1,\dots,N\}$, and $\vk(X,\chi^*)$ has row entries  $\vk(X,\chi^*)_{(i)} = k(\chi_i,\chi^*)$, $i\in\{1,\dots,N\}$, $\vk(\chi^*,X) = \vk(X,\chi^*)^{\T}$, $I$ is the identity matrix, and the noise variation $\sigma_w^2$ is another hyperparameter of the GP. The mean of the joint distribution is assumed to be zero here, since the GP is used to approximate the error dynamics $\delta f$ for which no prior knowledge is available. The hyperparameters $\Theta = (\sigma_w, \sigma_s. M)$ can be learned by solving a maximum log-likelihood problem
	 \begin{align}
	 	\Theta^* = \argmax_{\Theta} \left\{ \log \left( p(\delta f(X)|X,\Theta) \right) \right\}
	 \end{align}
	  through gradient ascent methods \cite{williams2006gaussian}.
	 
	 Given the joint probability distribution (\ref{eq:joint_pd}), the predictive mean and variance of the error function can be specified as
	 \begin{align}
	 	\Est{\delta f (\chi^*)} &= \vk(\chi^*,X)(K(X,X) + \sigma_w^2 I)^{-1} \delta f(X), \label{eq:df_mean} \\
	 	\Var{\delta f (\chi^*)} &= k(\chi^*,\chi^*) - \vk(\chi^*,X)(K(X,X) + \sigma_w^2 I)^{-1}\vk(X,\chi^*). \label{eq:df_var}
	 \end{align}
	 
	 Using the property of GP that the derivative of a GP is still a GP, the joint probability distribution of the observed output and the derivative of the predicted error function can be obtained as
	 \begin{align}
	 \begin{bmatrix}
	 \delta f(X) \\ \gradchi \delta f (\chi^*)   
	 \end{bmatrix}
	 \sim \mathcal{N} \left(0_{N}, \begin{bmatrix}
	 K(X,X) + \sigma_w^2 I & \gradchi \vk(X,\chi^*) \\
	 \gradchi \vk(\chi^*,X)  &  \gradcc k(\chi^*,\chi^*)
	 \end{bmatrix} \right). \label{eq:joint_pd_deriv}
	 \end{align}
	 
	The derivatives in (\ref{eq:joint_pd_deriv}) can be expressed analytically as follows
	\begin{align}
		\gradchi \vk(X,\chi^*)_{(i)} &= \gradchi \vk(\chi^*,X)_{(i)} = M  k(\chi_i,\chi^*)(\chi_i - \chi^*), \quad i\in\{1,\dots,N\},\\
		\gradcc k(\chi^*,\chi^*) &= \sigma_s^2 M.
	\end{align}
	
	Similar to the prediction process in error function $\delta f$, the predictive mean and variance of its state derivative is given by
	\begin{align}
	\Est{\gradchi \delta f (\chi^*)} &= \gradchi \vk(\chi^*,X) (K(X,X) + \sigma_w^2 I)^{-1} \delta f(X), \label{eq:df_deriv_mean} \\
	\Var{\gradchi \delta f (\chi^*)} &= \gradcc \vk(\chi^*,\chi^*)  - \gradchi \vk(\chi^*,X) (K(X,X) + \sigma_w^2 I)^{-1}\gradchi \vk(X,\chi^*). \label{eq:df_deriv_var}
	\end{align}
	
	The analytical expression of the state derivative obtained here can be useful in speeding up the computation process later on.
	
	\section{Nonlinear $\Hinf$ Control} \label{sec:hinf_control}
	
	In this section, we set up a nonlinear $\Hinf$ control problem to synthesize controllers that stabilize the system with guaranteed performance. 
	
	To this end, consider the general structure of the soft-constrained differential game associated with the disturbance attenuation problem in nonlinear $\Hinf$ control. A detailed explanation of the relationship between the $\Hinf$ control and the zero-sum differential game can be found in \cite{TamerBasar1995}.
	
	The dynamics of the differential game is described by
	\begin{align} \label{eqn:dynamic_system_minmax}
	\frac{\rd \vx(t)}{\rd t}  = F (\vx(t), \vu(t), \vv(t)), \quad \vx(t_0) = \vx_0, \quad t \ge 0,
	\end{align}
	where $\vv(t) \in U_2 \subset \Re^{q}$ denote the disturbance of the system and $U_2$ is a constraint set of the disturbance $\vv$, 
and the right-hand side of the dynamics in our problem is given by
\begin{align}
F (\vx, \vu, \vv) = \bar{f}(\vx, \vu) + \mu (\vx, \vu) + W(\vx, \vu) C(\vx,\vu) \vw,
\end{align}
where the time arguments are omitted for brevity, $\bar{f}(\vx, \vu)$ is the known nominal dynamics in (\ref{eq:dyn_ori}), and $\mu (\vx, \vu)$ is the mean of the GP-learned multivariate Gaussian distribution that approximates the unknown error dynamics $\delta f (\vx, \vu)$ such that $\delta f (\vx, \vu) \sim \mathcal{N} (\mu (\vx, \vu), \Sigma (\vx, \vu))$. Given the data set $\{X, \delta f(X)\}$, the mean and variance of the Gaussian distribution can be calculated as in (\ref{eq:df_mean}) and (\ref{eq:df_var}) respectively. In particular, $\mu (\vx, \vu) = \Est{\delta f (\chi)}$ and  $\Sigma (\vx, \vu) = \Var{\delta f (\chi)}$, where $\chi = (\vx, \vu)$.
The matrix $C(\vx,\vu) \in \Re^{n\times q}$ encodes the information of the structure of the noise that is known a priori. For instance, in many applications the noise comes in the same channel as the control, so for a control-affine system where $\bar{f}(\vx, \vu) = f_1(\vx) + f_2(\vx)\vu$, one can set $C(\vx,\vu) = f_2(\vx)$. If no prior structural information of the noise is assumed, then one can set $q = n$ and $C(\vx,\vu) = I$. The matrix
$W(\vx, \vu) = \Sigma (\vx, \vu)^{\frac{1}{2}}$ encodes the standard deviation information of the disturbance and this information is gained from the learning process. 
	
	The cost of the differential game is defined as 
	\begin{equation} \label{cost:eqn_dg}
	J_{\gamma}(\vu(\cdot), \vv(\cdot)) = \phi(\vx(t_f), t_f) + \int_{t_0}^{t_f} \mathcal{L}_{\gamma} (\vx(t), \vu(t), \vv(t))\, \text{d} t,
	\end{equation}
	where $t_0$ and $t_f$ are fixed initial and terminal time,  $\phi : \Re^{n} \times [t_0, \tf] \to \Re_{+}$ denotes the {terminal cost}, and
	\begin{equation}
		\mathcal{L}_{\gamma} (\vx(t), \vu(t), \vv(t)) = \mathcal{L} (\vx(t), \vu(t), \vv(t)) - \gamma^2 \vv(t)^{\T}\vv(t),
	\end{equation}
	where $\mathcal{L} : \Re^{n} \times \Re^{m} \times \Re^{q} \to \Re_{+}$ denotes the {running cost} of the performance index associated with the system (\ref{eqn:dynamic_system_minmax}), specified as
	\begin{equation} \label{cost:eqn}
	J(\vu(\cdot), \vv(\cdot)) = \phi(\vx(t_f), t_f) + \int_{t_0}^{t_f} \mathcal{L} (\vx(t), \vu(t), \vv(t))\, \text{d} t.
	\end{equation}
	
%
	
	In the differential game described by (\ref{eqn:dynamic_system_minmax}) and (\ref{cost:eqn_dg}), we wish to find  non-anticipative feedback strategy $\mu_{\vu} : [t_0, \tf] \times \Re^n \rightarrow U_1$ for the control and $\mu_{\vv} : [t_0, \tf] \times \Re^n \rightarrow U_2$ for the disturbance such that the corresponding cost function
		\begin{equation} \label{cost:eqn_dg_strategy}
		J_{\gamma}(\mu_{\vu}, \mu_{\vv}) = \phi(\vx(t_f), t_f) + \int_{t_0}^{t_f} \mathcal{L}_{\gamma} (\vx(t), \mu_{\vu}(t,\vx(t)), \mu_{\vv}(t, \vx(t)))\, \text{d} t
		\end{equation}
	 is minimized by the control while maximized by the disturbance. 
	 The function describing the minimax value of the cost function at $t_0$ and $\vx_0$ is given by
	 \begin{align} \label{perf_ind}
	 V(\vx_0, t_0)  
	 = \min_{\mu_{\vu}}  \max_{\mu_{\vv}} J_{\gamma}(\mu_{\vu}, \mu_{\vv}),
	 \end{align}
	 which is known as the value function.

	Standard regularity assumptions on the functions $F$, $\mathcal{L}$, and $\phi$ will be assumed throughout the paper to ensure existence and uniqueness of solutions of the differential equation (\ref{eqn:dynamic_system_minmax}).
	Accordingly,  it will be assumed that $\mu_{\vu}(t,\vx)$ and $\mu_{\vv}(t,\vx)$ are piecewise continuous functions in $t$ and Lipschitz continuous functions in $\vx$.

	
	It is further assumed that 
	both the control and disturbance have perfect knowledge of the dynamics of the system given by  (\ref{eqn:dynamic_system_minmax}), the constraint sets $U_1$ and $U_2$, the cost function, and the state $\vx$ at each current time. This is a standard assumption for differential games with complete information \cite{isaacs1999differential}.
	The Isaacs condition 
			\begin{equation} \label{eq:value}
			\hspace*{-3mm}
			\min\limits_{\vu } \max\limits_{\vv } \{\mathcal{L}(\vx,\vu,\vv) + \gradx V^{\T}F(\vx,\vu,\vv)\} = \max\limits_{\vv } \min\limits_{\vu } \{\mathcal{L}(\vx,\vu,\vv) + \gradx V^{\T}F(\vx,\vu,\vv)\},
			\end{equation}
	is assumed to hold  for  all $\vx,\vu,\vv$ and $t \ge 0$, which ensures that the $\min$ and $\max$ operators can be interchanged without affecting the outcome
	and thus ensures that the value of the game exists. 
	In (\ref{eq:value}), $ \gradx V(\vx,t)$ denotes the partial derivative of the value function $V(\vx,t)$ with respect to the state $\vx$.
	Notice that the Isaacs' condition holds when the controls are separable in both dynamics and cost~\cite{isaacs1999differential}, which is the case for many problems in practice.
	
	Under the assumption that the value of the game exists, and applying Theorem 4.14 in \cite{TamerBasar1995}, it can be shown that there exists some positive definite function $q_0(\vx)$, such that $\forall \vv:[t_0,\tf]\to \Re^{q}$ where $\vv(t) \in U_2 \subset \Re^{q}, t \in [t_0,\tf]$, and $\forall \vx_0 \in \Re^{n}$,
	\begin{align}
	J(\mu_{\vu}^*,\vv) \le \gamma^2 (\int_{t_0}^{\tf} \vv^{\T}\vv \text{d} t + q_0(\vx_0)),
	\end{align}
	that is, the optimal state feedback controller $\mu_{\vu}^*$ ensures a disturbance attenuation bound $\gamma$. 
	 For simplicity of the presentation, we henceforth use $\vu$ and $\vv$ to denote $\mu_{\vu}, \mu_{\vv}$, respectively. For instance, $V(\vx_0, t_0) = \min\limits_{{\vu}}  \max\limits_{{\vv}} J_{\gamma}({\vu}, {\vv})$.
	 

	\section{Game-Theoretic Differential Dynamic Programming}   \label{sec:GTDDP}
	
		Game-Theoretic Differential Dynamic Programming (GT-DDT) \cite{sun2018min} solves a differential game problem by 
		 utilizing the dynamic programming method and attempting to find  a  locally optimal control policy through the iterative improvement of a nominal control and  state trajectory. 
		The control update within each iteration of GT-DDT consists of two main steps. The first step is the optimal policy update, which can be shown to be a function  of the zero, first and second order of terms of the value function expansion along the nominal trajectory.
		The second step consists of the backward propagation of the local model of the value function.
%
%
%

	\subsection{Optimal Policy Update}
	
	The derivation of GT-DDP starts from the Hamilton-Jacobi-Isaacs (HJI) partial differential equation satisfied by the value function, which is derived from the dynamic programming principle \cite{basar1999dynamic, isaacs1999differential}.
	In particular, the value of the game in (\ref{perf_ind}) satisfies
	\begin{align} \label{eqn:HJI_minmax}
	- \frac{\partial V(\vx,t)}{ \partial t} &= \min_{\vu}  \max_{\vv} Q(\vx,\vu,\vv,t),
	\end{align}
	where 
	\begin{align} \label{eqn:Q_function}
	Q(\vx,\vu,\vv,t) &= \mathcal{L}_{\gamma}(\vx,\vu,\vv)  + \gradx V(\vx, t)^{\T} F(\vx,\vu,\vv),
	\end{align}
	with the boundary condition at time $t = \tf$ 
	\begin{equation}  \label{eqn:HJB_final_cond}
	V(\vx(\tf), \tf) = \phi(\vx(\tf), \tf).
	\end{equation}
	
	The optimal control and disturbance update laws $\delta \vu^{*}$, $\delta \vv^{*}$ can be derived from second order expansion of the terms in  both sides of equation (\ref{eqn:HJI_minmax}) about the nominal state, control and disturbance  trajectories $({\bar{\vx}}, {\bar{\vu}}, {\bar{\vv}})$.  
	Specifically, 
	  \begin{align}
	&  -  \frac{\text{d} V}{\text{d} t}    - \delta \vx^{\T}   \frac{\text{d} (\gradx V)}{\text{d} t}  - \frac{1}{2} \delta \vx^{\T}  \frac{\text{d} (\gradxx V)}{\text{d} t} \delta \vx \nonumber \\
	& =  \min_{\delta \vu} \max_{\delta \vv} Q(\bar{\vx} + \delta \vx,\bar{\vu} + \delta \vu,\bar{\vv} + \delta \vv,t) \nonumber \\
	& = \min_{\delta \vu} \max_{\delta \vv}   \left\{ \mathcal{L} +  \delta \vx^{\T} {Q_{\vx}} + \delta \vu^{\T} {Q_{\vu}} + \delta \vv^{\T} {Q_{\vv}}  + \frac{1}{2} \left[ \begin{array}{c} \delta \vx \\  \delta \vu  \\  \delta \vv  \end{array} \right]^{\T}  \left[ \begin{array}{ccc}
	Q_{\vx \vx}  &  Q_{\vx \vu}  & Q_{\vx \vv}  \\
	Q_{\vu \vx}  & Q_{\vu \vu}  & Q_{\vu \vv} \\
	Q_{\vv \vx}  & Q_{\vv \vu}  &  Q_{\vv \vv}
	\end{array} \right]  \left[ \begin{array}{c} \delta \vx \\  \delta \vu  \\  \delta \vv  \end{array} \right] \right\}, \label{eqn:MinMaxContinuous_QFunction}
	\end{align}
	where
	\begin{subequations} \label{eq:Q_func_cont}
		\begin{alignat}{1}
	&Q_{\vx}  =    \gradx F^{\T} \gradx V  + \gradx \mathcal{L},  \\
	&Q_{\vu}  =    \gradu F^{\T} \gradx V  + \gradu \mathcal{L} , \\
	& Q_{\vv} =  \gradv F^{\T} \gradx V  + \gradv \mathcal{L},  \\
	& Q_{\vx \vx}= \gradxx \mathcal{L} + 2 \gradxx  V \gradx  F, \\
	& Q_{\vu \vu}  = \graduu \mathcal{L}, \\
	& Q_{\vv \vv} =  \gradvv \mathcal{L}, \\
	& Q_{\vu \vx} =   \gradu  F^{\T} \gradxx V   +  \gradux \mathcal{L} , \\
	& Q_{\vv \vx} = \gradv F^{\T} \gradxx V   +  \gradvx \mathcal{L},  \\
	& Q_{\vu \vv} =  \graduv  \mathcal{L}. \\
	& Q_{\vx \vu} =  Q_{\vu \vx}^{\T}, \quad  Q_{\vx \vv} =  Q_{\vv \vx}^{\T}, \quad  Q_{\vv \vu} =  Q_{\vu \vv}^{\T}.
		\end{alignat}
	\end{subequations}

The arguments of the functions $Q, V, F$, etc, are omitted for brevity, and the equations in (\ref{eq:Q_func_cont}) are evaluated along the nominal trajectory $(\bar{\vx},\bar{\vu}, \bar{\vv})$.
	

	The optimal control $ \delta \vu^{*} $ and disturbance $ \delta \vv^{*} $ must be chosen to minimize and maximize the right-hand side of (\ref{eqn:MinMaxContinuous_QFunction})  respectively. By computing the gradients of (\ref{eqn:MinMaxContinuous_QFunction}) with respect to $\delta \vu$ and $ \delta \vv$, respectively, and set them equal to zero, one obtain
	\begin{subequations}  \label{eqn:OptimalControl_uv}
	\begin{alignat}{1}
	\delta \vu^{*} &= - Q_{\vu \vu}^{-1} \bigg( Q_{\vu \vx} \delta \vx  + Q_{ \vu \vv} \delta \vv^{*}    + Q_{\vu}  \bigg)  , \label{eqn:OptimalControl_u} \\
	\delta \vv^{*} &= - Q_{\vv \vv}^{-1} \bigg( Q_{\vv \vx} \delta \vx  + Q_{ \vv \vu} \delta \vu^{*}    + Q_{\vv}  \bigg) , \label{eqn:OptimalControl_v}
	\end{alignat}
	\end{subequations}
	where $Q_{ \vv \vu} = Q_{ \vu \vv}^{\T}$.	
	
Notice that $\delta \vu^{*}$ and $\delta \vv^{*}$ show up in the right-hand side of (\ref{eqn:OptimalControl_uv}), and this can be further derived to obtain the
explicit expressions  for  $  \delta \vu^{*}  $   and  $ \delta \vv^{*} $ as
\begin{subequations}  \label{eqn:OptimalControl_uv_mod}
	\begin{alignat}{1}
		\delta \vu^{*}    &   = \vl_{\vu}   + \vK_{\vu}   \delta \vx \\
		\delta \vv^{*}    &   = \vl_{\vv}   + \vK_{\vv}   \delta \vx,
	\end{alignat}
\end{subequations}
	where the feed-forward gains $\vl_{\vv}  ,\vl_{\vu}      $ and feedback gains $\vK_{\vv}  ,\vK_{\vu}    $  are given by
	\begin{equation} 
	\vl_{\vu}   =  -   \bigg( Q_{\vu \vu}    -  Q_{\vu \vv} Q_{\vv \vv}^{-1} Q_{\vv \vu}  \bigg)^{-1}    \bigg( Q_{\vu}  -  Q_{\vu \vv} Q_{\vv \vv}^{-1}Q_{\vv}  \bigg), \label{eqn:gain_lu} \\
	\end{equation}
	\begin{equation} \label{eqn:gain_lv}
	\vl_{\vv}   =   -   \bigg(  Q_{\vv \vv}   -    Q_{\vv \vu} Q_{\vu \vu}^{-1} Q_{\vu \vv}  \bigg)^{-1}  \bigg( Q_{\vv}  -   Q_{\vv \vu} Q_{\vu \vu}^{-1}   Q_{\vu}   \bigg),
	\end{equation}
	\begin{equation}  \label{eqn:gain_Lu}
	\vK_{\vu}   =  - \bigg( Q_{\vu \vu}  -   Q_{\vu \vv} Q_{\vv \vv}^{-1} Q_{\vv \vu}  \bigg)^{-1}   \bigg(  Q_{\vu \vx}  - Q_{\vu \vv} Q_{\vv \vv}^{-1} Q_{\vv \vx} \bigg),
	\end{equation}
	\begin{equation} \label{eqn:gain_Lv}
	\vK_{\vv}   = -\bigg(  Q_{\vv \vv}  -   Q_{\vv \vu} Q_{\vu \vu}^{-1} Q_{\vu \vv}  \bigg)^{-1}  \bigg( Q_{\vv \vx}  -  Q_{\vv \vu} Q_{\vu \vu}^{-1}Q_{\vu \vx} \bigg).
	\end{equation}
	
%

	\subsection{Value Function Propagation}
	
	The update law for the second order local model of the value function
	$(V, \gradx V, \gradxx V)$ can be derived by substituting the optimal control (\ref{eqn:OptimalControl_u}) and disturbance (\ref{eqn:OptimalControl_v}) to the HJI equation (\ref{eqn:HJI_minmax}) and collecting corresponding terms, which yields
		\begin{subequations} \label{eqn:riccati_eqs}
			\begin{alignat}{1}
			-  \frac{\text{d} V}{\text{d} t}   &   =    \mathcal{L}  + \vl_{\vu} ^{\T}Q_{\vu }+  \vl_{\vv} ^{\T} Q_{\vv} +  \frac{1}{2} \vl_{\vu}   Q_{\vu \vu}  \vl_{\vu}   +    \vl_{\vu} ^{\T} Q_{\vu \vv}\vl_{\vv} +    \frac{1}{2}\vl_{\vv} ^{\T} Q_{\vv \vv}\vl_{\vv} ,   \\
			- \frac{\text{d} (\gradx V)}{\text{d} t} & =   Q_{\vx} +  \vK_{\vu}^{\T} Q_{\vu } + \vK_{\vv}^{\T} Q_{\vv}  + Q_{\vu \vx}^{\T} \vl_{\vu}+ Q_{\vv \vx}^{\T} \vl_{\vv}  + \vK_{\vu}^{\T} Q_{\vu \vu} \vl_{\vu}  +  \vK_{\vu}^{\T} Q_{\vu \vv} \vl_{\vv}  \nonumber \\
			& +  \vK_{\vv}^{\T} Q_{\vv \vu} \vl_{\vu}  +  \vK_{\vv}^{\T} Q_{\vv \vv} \vl_{\vv} , \\
			- \frac{\text{d} (\gradxx V)}{\text{d} t} & =        \vK_{\vu} ^{\T}Q_{\vu \vx}  +  Q_{\vu \vx}^{\T}\vK_{\vu} +  \vK_{\vv} ^{\T}  Q_{\vv \vx}  +  Q_{\vv \vx}^{\T} \vK_{\vv}   +   \vK_{\vv}^{\T}  Q_{\vv \vu} \vK_{\vu} +   \vK_{\vu}^{\T}  Q_{\vu \vv} \vK_{\vv} \nonumber \\
			& +   \vK_{\vu}^{\T}  Q_{\vu \vu} \vK_{\vu}  +   \vK_{\vv}^{\T} Q_{\vv \vv} \vK_{\vv}
			+    Q_{\vx \vx},
			\end{alignat}
		\end{subequations}
	under the boundary condition at time $t = \tf$ 
	\begin{subequations} \label{eq:boundary_cond_riccati}
		\begin{alignat}{1}
		V(\tf) &= \phi(\bar{\vx}(\tf), \tf) , \\
		\gradx V(\tf) &= \phi_{\vx} (\bar{\vx}(\tf), \tf) , \\
		\gradxx V(\tf) &= \phi_{\vx \vx} (\bar{\vx}(\tf), \tf).
		\end{alignat}
	\end{subequations}

\subsection{Local Control Policy in Application} \label{subsec:policy}
Upon convergence of the GT-DDP iteration, it finds a locally optimal trajectory $\vx^{\star}$
along with the corresponding optimal control trajectory $\vu^{\star}$ and disturbance trajectory $\vv^{\star}$.
A feedback controller is normally desired in real-world applications in order to cope with modeling errors or unknown disturbances. Via the GT-DDP algorithm, a local policy along the optimized trajectory is readily available 
\begin{equation} \label{eqn:feedback_ctrl}
	\vu(\vx) = \vu^{\star} + \vK_{\vu} (\vx - \vx^{\star}),
\end{equation}
where $\vK_{\vu}$ is a time dependent gain matrix given by (\ref{eqn:gain_Lu}).

	\section{Numerical Examples} \label{sec:Simulation_Results}
	
	In this section, we  apply our algorithm to a quadcopter system to test its performance.


The dynamic model of the quadcopter has 16 states, including the position ($\vr = (x, y, z)^{\T}$), the Euler angles ($\Phi = (\phi, \theta, \psi)^{\T}$), the velocity ($\dot{\vr} = (\dot{x}, \dot{y}, \dot{z})^{\T}$), the body angular velocity ($\dot{\Phi} = (p, q, r)^{\T}$) and the motor speeds ($\Omega = (\omega_1, \omega_2, \omega_3, \omega_4)^{\T}$). The nominal model of the quadcopter is given as follows:

\begin{align} 
\bar{f}(\vx(t), \vu(t))  = f(\vx) + G \vu, \label{eqn:quadcopter_dyn}
\end{align}

\begin{figure}[htb]
	\centering
	\includegraphics[width=0.5\textwidth]{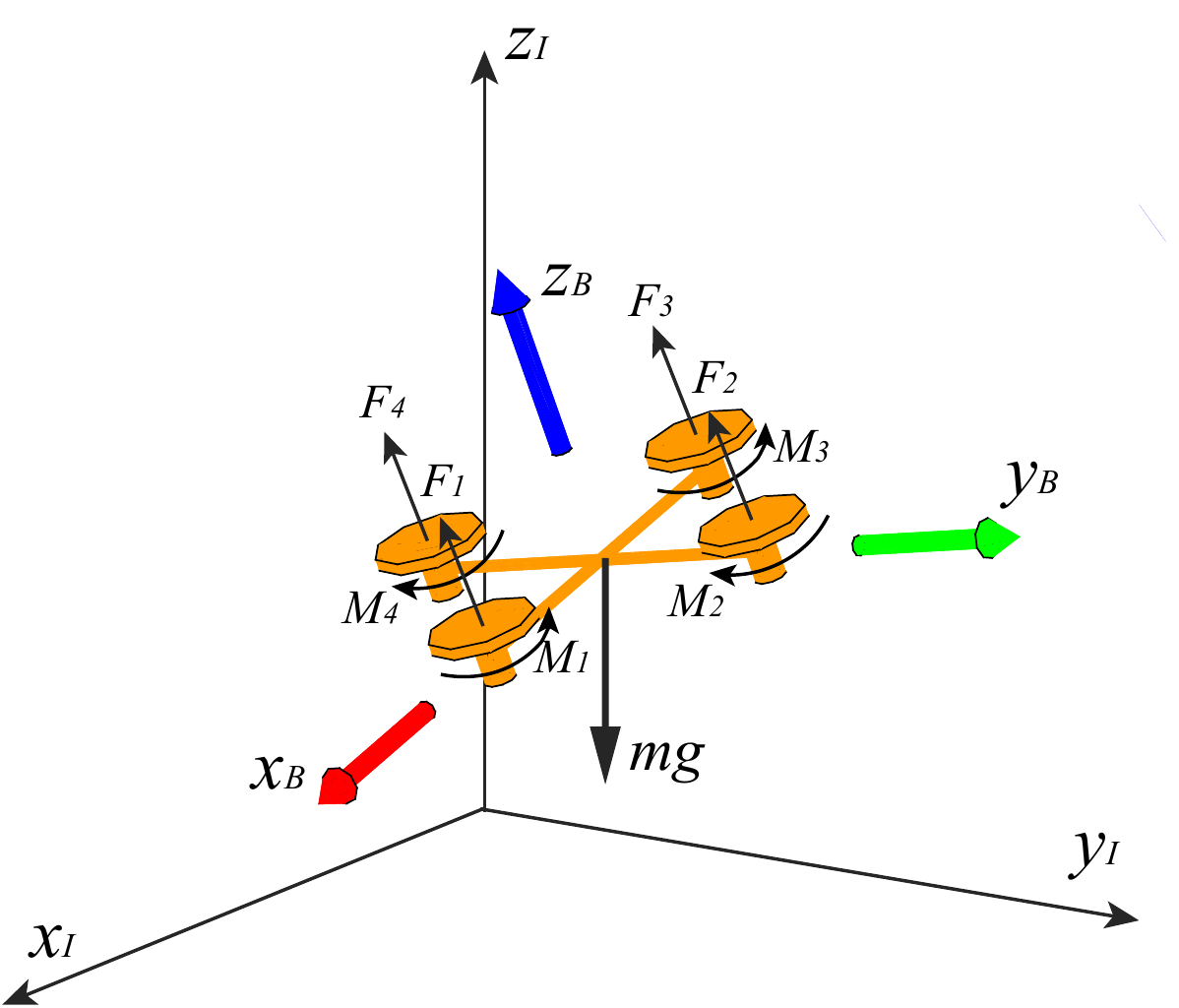}
	\caption{Coordinate systems and forces/moments acting on a quadcopter.}
	\label{fig:quad}
\end{figure}
where $\vx = [\vr, \Phi, \dot{\vr}, \dot{\Phi}, \Omega]^{\T} \in \Re^{16}$, 
\begin{align} \label{eqn:quad_f}
f(\vx) = 
&\begin{pmatrix}
\dot{x} \\
\dot{y}  \\
\dot{z} \\
\begin{bmatrix}
\cos \theta & 0 & -\cos \phi \sin \theta \\
0 & 1 & \sin \phi \\
\sin \theta & 0 & \cos \phi \cos \theta
\end{bmatrix}^{-1}
\begin{pmatrix}
p \\ q \\ r
\end{pmatrix} \\[15pt]
\begin{pmatrix}
0 \\ 0 \\ g
\end{pmatrix} + 
R \begin{pmatrix}
0 \\ 0 \\ \frac{1}{m} \sum F_i
\end{pmatrix} \\[15pt]
I^{-1} \left[ \begin{pmatrix}
L(F_2 - F_4) \\ L(F_3 - F_1) \\ M_1 - M_2 + M_3 - M_4
\end{pmatrix} - 
\begin{pmatrix}
p \\ q \\ r
\end{pmatrix} \times I
\begin{pmatrix}
p \\ q \\ r
\end{pmatrix} \right] \\[15pt]
k_m \Bigg[ \begin{pmatrix}
\omega_h \\ \omega_h \\ \omega_h \\ \omega_h 
\end{pmatrix} - \begin{pmatrix}
\omega_1  \\ \omega_2  \\ \omega_3 \\ \omega_4
\end{pmatrix} \Bigg]
\end{pmatrix},
\end{align}
where $\omega_h = \sqrt{\frac{mg}{4 k_F}}$ is the motor speed required for hovering, $m$ is the mass of the quadcopter, $g$ is gravitational acceleration, $L$ is the length of the moment arm from the body of the quadcopter to the motor, $I$ is the mass moment of inertia matrix for the quadcopter in the body frame, and $k_m$ is a motor constant. The forces and moments $F_i$ and $M_i, i = 1, \dots, 4$ are given by 
\begin{align}
F_i &= k_F \omega_i^2, \\
M_i &= k_M \omega_i^2.
\end{align}
The motor constants $k_m, k_F$ and $k_M$ are given by $k_m = \frac{1}{20}, k_F = 6.11 \times 10^{-8} \frac{N}{rpm^2}, k_M = 1.5 \times 10^{-9} \frac{N \cdot m}{rpm^2}$.
$R$ is the rotation matrix between the body and inertial frames, which can be calculated by 
\begin{align}
R = \begin{bmatrix}
c \psi c \theta - s \phi s \psi s \theta & - c \phi s \psi & c \psi s \theta + c \theta s \phi s \psi \\
c \theta s \psi + c \psi s \phi s \theta & c \phi c \psi & s \psi s \theta - c \psi c \theta s \phi \\
- c \phi s \theta & s \phi & c \phi c \theta
\end{bmatrix},
\end{align}
where `$s$' and `$c$' in the previous matrix are short for `$\sin$' and `$\cos$', respectively.

$G$ is a $16 \times 4$ constant matrix:
\begin{align} \label{eqn:quad_G}
G = k_m \begin{bmatrix}
0_{1,1} & 0_{1,2} & 0_{1,3} & 0_{1,4} \\
\vdots & \vdots & \vdots & \vdots \\
0_{12,1} & 0_{12,2} & 0_{12,3} & 0_{12,4} \\
1 & 0 & -1 & 1 \\
1 & 1 & 0 & -1 \\
1 & 0 & 1 & 1 \\
1 & -1 & 0 & -1
\end{bmatrix},
\end{align}
and $\vu = (u_1, u_2, u_3, u_4)^{\T} \in \Re^4$ is the control vector, where $u_1$ results in a net force along the $z_B$ axis and $u_2, u_3, u_4$ generate roll, pitch and yaw moments, respectively, whereas $\vv \in \Re^4$ denotes the disturbance.

The dynamics of the  soft-constrained differential game is
\begin{align}
	\frac{\rd \vx}{\rd t} = \bar{f}(\vx, \vu) + \mu (\vx, \vu) + W(\vx, \vu) C(\vx,\vu) \vw,
\end{align}
where $C(\vx,\vu) = G$, $\mu (\vx, \vu)$ and $W(\vx,\vu)$ are learned by training a Gaussian Process. The data points are obtained through the simulation of dynamics
\begin{align}
  \rd \vx =	\tilde{f}(\vx, \vu) \rd t + G \rd \omega, \quad \vx(t_0) = \vx_0, \quad \rd  \omega \sim \mathcal{N}(0,\Sigma_w), \label{eq:dyn_qr_stoch}
\end{align}
where $\tilde{f}$ has the same structure as $\bar{f}$ but varied parameters in $I$ and $L$.

\begin{figure}[htb]
	\centering
		\includegraphics[width=0.40\textwidth]{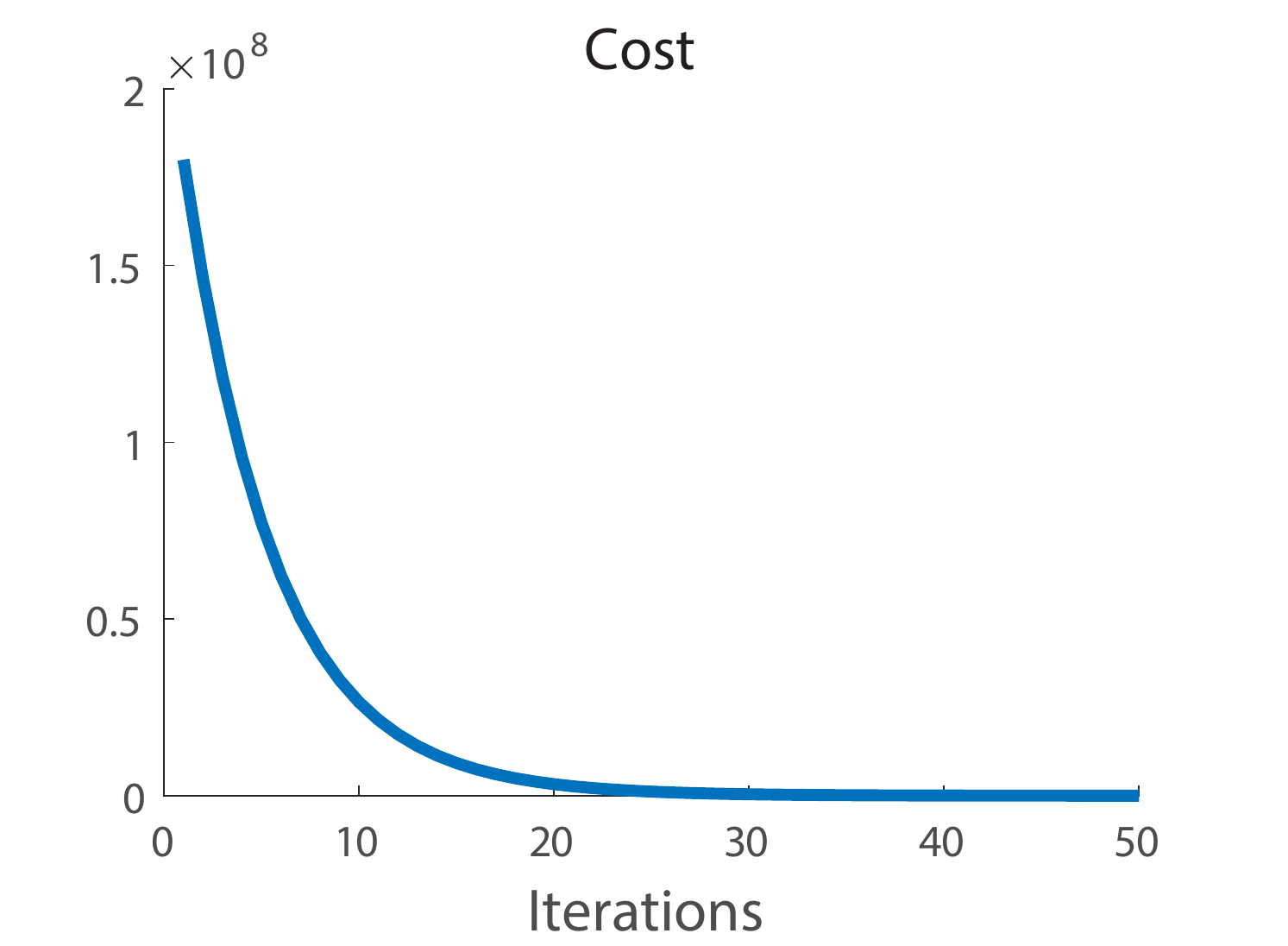}
		\caption{Cost per iteration.}
		\label{fig:cost_quad_minimax}
\end{figure}

The corresponding cost function of the game is defined as
\begin{equation}
J_{\gamma} = \int_{0}^{\tf} [ (\vx - \vx_f)^{\T} Q (\vx - \vx_f) + \vu^{\T} R_{\vu} \vu -  \gamma^2 \vv^{\T} \vv ] + (\vx - \vx_f)^{\T} Q_f (\vx - \vx_f),
\end{equation}
 where $\vx_f \in \Re^{16}$ denotes the desired terminal states. 
 The goal of this task is to steer the quadcopter to $\vx_f$, where $\vx_{f1}  = 3$, $\vx_{f2}  = 5$, $\vx_{f3}  = 1$, $\vx_{f6}  = \pi$ and $\vx_{fi} = 0$ otherwise.
 In the simulations, we set 
and
\begin{align}
	[Q_f]_{i, i} =  \begin{cases}
		10^7, \quad i = 1\text{--}3;\\
		10^6, \quad i = 4\text{--}9;\\
		10^5, \quad i = 10\text{--}12;\\
		0, \quad \text{otherwise},
	\end{cases}
\end{align}
and all the off-diagonal terms are assigned to zero.
We also chose
$Q = 0.00001 Q_f$, $R_{\vu} = 0.0001 I$ and $\gamma = 0.05$.
%
The cost per iteration of the GT-DDP algorithm is shown in Figure~\ref{fig:cost_quad_minimax}.


\begin{figure}[htb]
	\centering
	\includegraphics[width=0.80\textwidth]{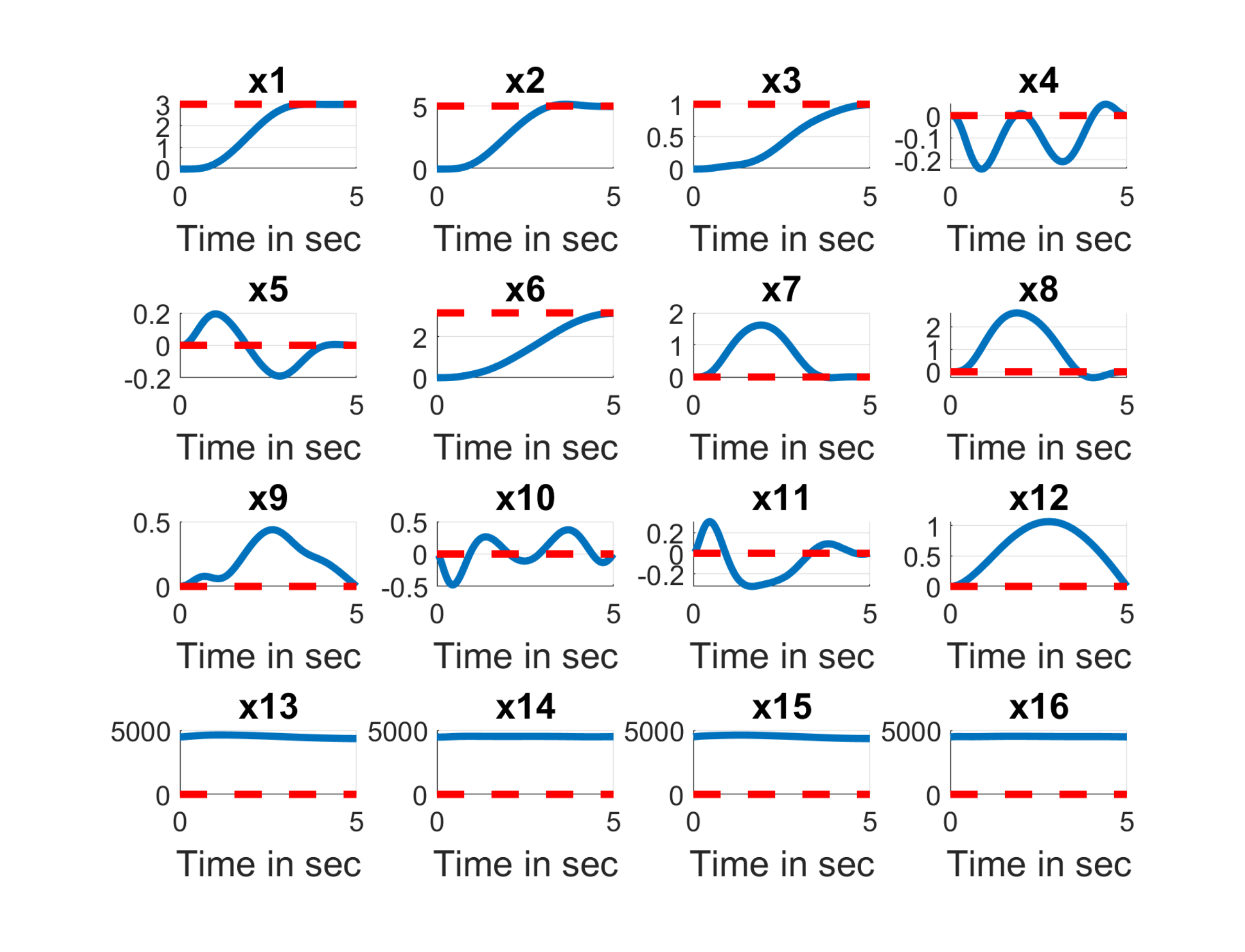}
	\caption{Mean of the state trajectories in 100 simulations in blue and corresponding goal states in dashed red.}
	\label{fig:optimal_traj_quad_minimax}
\end{figure}

The feedback local policy (\ref{eqn:feedback_ctrl}) is then applied to the dynamics (\ref{eq:dyn_qr_stoch}) under stochastic disturbance. The mean of state trajectories in 100 simulations is shown in Figure~\ref{fig:optimal_traj_quad_minimax}. It can be seen that the feedback control guide the state to its desired goal despite the presence of stochastic disturbances. 

\section{Conclusion} \label{sec:Conclusion}

In this paper, we have introduced an approach that combines machine learning with nonlinear $\Hinf$ control to design a learning-based controller that takes into account the learned model dynamics and disturbance estimate while achieving performance guarantee of the system. The error dynamics and disturbance estimate are learned through the utilization of the Gaussian Process (GP) in order to formulate the soft-constrained differential game associated with the disturbance attenuation nonlinear $\Hinf$ control problem. The differential game is then solved through the Game-Theoretic Differential Dynamic Programming (GT-DDP) method to obtain the optimal state-feedback control that guarantees performance despite model uncertainties. 
The GT-DDP method is derived by taking the Taylor series expansion of the Hamilton-Jacobi-Isaacs (HJI) equation associated with the differential game around a nominal trajectory to obtain the update law of the control and disturbance, as well as the backward propagation equations of the approximation terms of the value function.
The approach we presented is tested on a nonlinear control problem, namely, a quadcopter flight steering.
The example showcases the ability of our controller to handle relatively high degree-of-freedom systems and system disturbances. Future possible extension of this research includes applying this method to problems with stochastic dynamics and state/control constraints, which will be suitable for a variety of applications including bio-mechanical systems and stochastic models in finance. 

	\vspace*{2ex}

\bibliographystyle{aiaa}
\bibliography{wei}

\end{document}